\documentstyle[seceq,twocolumn]{jpsj}
\def\v#1{\mib #1}
\def\sbox#1{\mbox{\scriptsize #1}}
\def\deltac{\delta_{\sbox{c}}}
\def\jc{j_{\sbox{c}}}

\def\runtitle{Ground State of XXZ Model with Dimerization and Quadrumerization}
\def\runauthor{Wei {\sc Chen}, Kazuo {\sc Hida}}

\title
{
Ground State Phase Diagram of the One Dimensional $\v{S}=1/2$ XXZ Model with Dimerization and Quadrumerization
}

\author
{
Wei {\sc Chen}\footnote{E-mail: chenwei@riron.ged.saitama-u.ac.jp} and Kazuo {\sc Hida}\footnote{E-mail: hida@riron.ged.saitama-u.ac.jp}
}

\inst
{
Department of Physics, Saitama University, Urawa, Saitama 338-8570
}

\recdate
{
\today
}

\abst
{
The one dimensional $S=1/2$ XXZ model with dimerization $(1-j)$ and quadrumerization $\delta$ is studied by the numerical exact diagonalization of finite size systems. Using the conformal field theory and the level spectroscopy method, we calculate the ground state phase diagram with XY-like anisotropy $\Delta$ $(0<\Delta<1)$. The ground states of this model contain the Haldane state, $S=1$ dimer state, $S=1$ large-{\it {D}} state and $S=1/2$ dimer state as limiting cases. The $\Delta$ and $\delta$-dependence of the critical exponent $\nu$ of the energy gap is determined from the conformal dimensions of excited states.
}

\kword
{
exact diagonalization, conformal field theory, dimer phase, Haldane phase, large-{\it {D}} phase
}

\begin{document}
\sloppy
\maketitle

\section{Introduction}
Recently, the one dimensional $S=1/2$ Heisenberg chains with a variety of spatial structures have attracted a great deal of attention. Although the uniform $S=1/2$ XXZ model can be solved by the Bethe's hypothesis\cite{ya} and is known to have a gapless ground state for the XY-like anisotropy, the various spatial structures such as the spin-Peierls dimerization and ladder structure can drive this ground state into the spin gap states.

In our previous work,\cite{wk} the phase diagram of the $S=1/2$ isotropic Heisenberg model with dimerization and quadrumerization was studied by the exact diagonalization and phenomenolozical renormalization group method. The XY case is also studied analytically. However, the phase diagrams in these two limiting cases are quite different. Therefore it is worthwhile to investigate the crossover between these two limiting cases.

This paper is organized as follows. In the next section, the model Hamiltonian is defined. The numerical results are presented in {\S}3. Analyzing the exact diagonalization data by the conformal field theory and the level spectroscopy method, we calculate the exponent $\nu$ characterizing the opening of the energy gap. The last section is devoted to summary and discussion.
\section{Model Hamiltonian}
The Hamiltonian of the one dimensional dimerized and quadrumerized $S=1/2$ XXZ chain is given by
\begin{eqnarray}
{\cal H} &=& \sum_{l=1}^{2N}  j(S_{2l-1}^{x}S_{2l}^{x}+S_{2l-1}^{y}S_{2l}^{y}+\Delta S_{2l-1}^{z}S_{2l}^{z})   \nonumber \\
& +& \sum_{l=1}^{2N}  (1+(-1)^{l-1} \delta)(S_{2l}^{x}S_{2l+1}^{x}+S_{2l}^{y}S_{2l+1}^{y} \nonumber \\
&+&\Delta S_{2l}^{z}S_{2l+1}^{z}),
\end{eqnarray}
where $1-j(-\infty \le j \le \infty)$ and $\delta(-1 \le \delta \le 1)$ represent the degree of dimerization and quadrumerization, respectively. The anisotropy parameter is denoted by $\Delta$. In our previous work\cite{wk}, we calculated the phase diagrams for the cases $\Delta=1$ (isotropic case) and 0 (XY case). In the present work, we concentrate on the XY-like  regime $ 0 < \Delta < 1$.

\section{Numerical Results}
\subsection{Phase diagram for $0 < \Delta < 1$}
In order to determine the phase boundary  with high accuracy, we use the twisted boundary method of Kitazawa and Nomura \cite{ak,kn}. The Hamiltonian is numerically diagonalized to calculate the two low lying energy levels with the twisted boundary condition $S^{x}_{4N+1}=-S^{x}_1,$ $ S^{y}_{4N+1}=-S^{y}_1,$ $ S^{z}_{4N+1}=S^{z}_1$ for $4N=12, 16, 20$ and 24.

From the result of the isotropic and XY cases\cite{wk}, we expect two different kinds of ground states. For small values of $j$ and $\delta$, the ground state is the Haldane-like state with valence bond solid (VBS) structure. Under the twisted boundary condition, the eigenvalues of the space inversion $P$ and the spin reversal $T$ are all equal to $-1$\cite{ak,kn}. As  $j$ and/or $\delta$ increases, the phase transition takes place into the dimer-like state for which $P=1$ and $T=1$. We make use of $P$ and $T$ eigenvalues to distinguish the two phases. For fixed $\delta$, the energies of the two states vary with $j$. For small $\mid j \mid $, the energy of the Haldane state is lower than that of the dimer state. As $\mid j \mid$ increases, the latter becomes lower than the former. The two levels cross at one point which gives the finite size transition point $j=\jc(N)$. Fig. \ref{fig1} shows the $j$-dependence of the two lowest levels for $N=24, \Delta=0.5, \delta=0.4$ and $j>0$. We extrapolate the critical point as $\jc(N)=\jc(\infty)+aN^{-2}+bN^{-4}.$

In Fig. \ref{fig2}, the ground state phase boundaries are represented by the solid line, $\triangle$, $\Box$ and $\circ$ for $\Delta =0, 0.5, 0.7$ and 1, respectively. For $0 \leq \Delta < 1$, the phase boundaries are closed at the negative finite value of $j = \jc(\delta)$. This is in contrast to the isotropic case in which the Haldane phase remains stable even in the limit $j \rightarrow -\infty$ as far as $\mid \delta \mid < \deltac \simeq 0.25$. Here the critical value $\deltac$ corresponds to the critical dimerization of the Haldane-dimer transition of the $S=1$ Heisenberg chain.\cite{tk,kno} This behavior can be understood in the similar way as the XY case\cite{wk} and  the ground state for $j < \jc(\delta)$ corresponds to the large-$D$ phase of the $S=1$ Heisenberg chain for $\mid \delta \mid < \deltac$. It should be noted that  there is no critical point between the large-{\it{D}} phase and the dimer phase as in the case of the $S=1$ dimerized Heisenberg chain.\cite{tt}

For $j \rightarrow 0$ and $\delta \rightarrow 1$, the spins connected by the $1+\delta$-bonds form strong singlet pairs. The effective coupling $j_{\sbox {eff}}$ and effective anisotropy $\Delta_{\sbox {eff}}$ between the spins $\v {S}_{4l+1}$ and $\v {S}_{4l+4}$ can be calculated by the perturbation theory as,
\begin{equation}
j_{\sbox {eff}}=\frac{j^{2}}{(1+\delta)(1+\Delta)},
\end{equation}
\begin{equation}
\Delta_{\sbox {eff}}=\frac{\Delta^{2}(\Delta+1)}{2}.
\end{equation}
Therefore the effective Hamiltonian is given by
\begin{eqnarray}
{\cal H_{\sbox {eff}}} &=& j_{\sbox {eff}}\sum_{l=0}^{N} (S_{4l+1}^{x}S_{4l+4}^{x}+S_{4l+1}^{y}S_{4l+4}^{y}+\Delta_{\sbox {eff}} S_{4l+1}^{z}S_{4l+4}^{z}) \nonumber \\
&+& (1-\delta)\sum_{l=1}^{N}(S_{4l}^{x}S_{4l+1}^{x}+S_{4l}^{y}S_{4l+1}^{y}+\Delta S_{4l}^{z}S_{4l+1}^{z}).
\end{eqnarray}
We have also diagonalized this effective model to check the phase boundary for small $j$. The phase boundary obtained from the effective model is shown by $\times$ in Fig. \ref{fig2} for the case $\Delta=0.5$.
\subsection{Critical exponent $\nu$ of the energy gap}
At the critical points, we calculate the energy with fixed wave number $k$ by Lanczos exact diagonalization method under the periodic boundary condition for the magnetization $M^z=0$ and 1 where $M^{z}=\displaystyle\sum_{l=1}^{4N}S_{l}^{z}$. The system sizes are $4N=12, 16, 20$ and 24. The ground state has $M^{z}=0$ and $k=0$.

It is known that the finite size correction to the ground state energy is related with the central charge $c$ and the spin wave velocity $v_{\sbox{s}}$ as follows, \cite{ca,HW,IA}
\begin{equation}
\label{eq1}
\frac{1}{N}E_{\sbox{g}}(N) \cong \varepsilon_{\infty}-\frac{\pi cv_{\sbox{s}}}{6N^{2}},
\end{equation}
\begin{equation}
\label{eq2}
v_{\sbox{s}}=\lim_{N \rightarrow \infty}\frac{N}{2\pi}[E_{k_{1}}(N)-E_{\sbox{g}}(N)],
\end{equation}
where $E_{\sbox{g}}(N)$ is the ground state energy, $E_{k_{1}}(N)$, the energy of the excited state with $k_{1}=\frac{2\pi}{N}$ and $M^z=0$, $c$, the central charge and $\varepsilon_{\infty}$, the ground state energy per unit cell in the thermodynamic limit. After appropriate extrapolation to $N \rightarrow \infty$,\cite{wkh} we have checked that the central charge $c$ is close to unity on the phase boundary. Therefore, the present model can be described by the Gaussian model on the critical line,
\begin{equation}
{\cal H_{\sbox{G}}}= \frac{1}{2\pi} \int {dx \Big[ v_{\sbox{s}}K (\pi\Pi)^{2}+\frac{v_{\sbox{s}}}{K}(\frac{\partial \phi}{\partial x})^{2} \Big]},
\end{equation}
where $\phi$ is the boson field defined in the interval $0 \leq \phi < \sqrt{2}\pi$ and $\Pi$ is the momentum density conjugate to $\phi$ which satisfies $[\phi(x),\Pi(x')]=i\delta(x-x')$.
The deviation from the critical point is described by the term $\cos \sqrt{2} \phi$ and the low energy properties of our model can be described by the one-dimensional quantum sine-Gordon theory near the critical line,
\begin{eqnarray}
{\cal H_{\sbox{SG}}}&=& \frac{1}{2\pi} \int {dx \Big[ v_{\sbox{s}}K (\pi\Pi)^{2}+\frac{v_{\sbox{s}}}{K}(\frac{\partial \phi}{\partial x})^{2} \Big]}\nonumber\\
&+& \frac{y_{1}v_{\sbox{s}}}{2\pi a^{2}} \int{dx \cos \sqrt{2}\phi}.
\end{eqnarray}
The scaling dimensions $x_n$ of the operators are related with the energy eigenvalues $E_{n}(N,M^{z})$ of the corresponding excited states as $x_n =\lim_{N \rightarrow \infty} x_n(N)$ where,
\begin{equation}
\label{eq3}
x_{n}(N)=\frac{N}{2\pi v_{\sbox{s}}}[E_{n}(N,M^{z})-E_{\sbox{g}}(N)].
\end{equation}
We can identify the correspondence between the operators in boson representation and the eigenstates of spin chains by comparing their symmetry properties.\cite{kn,nomura} Let us denote the scaling dimensions of the operators $\sqrt{2}{\mbox {cos}}\sqrt{2}\phi$ and $\sqrt{2}{\mbox {sin}}\sqrt{2}\phi$ by $x_{1}$ and $x_{2}$, respectively. Both of them should be equal to $K/2$ in the thermodynamic limit as determined from their correlation functions.  Numerically, these exponents are calculated using Eq.(\ref{eq3}) for $M^{z}=0$, $P=1$, $k=0 [x_1]$ and  $M^{z}=0$, $P=-1$, $k=0 [x_2]$.  Thus the value of $K$ can be determined from these values. Actually, to reduce the finite size correction to $O(1/N^{2})$, it is more convenient to use the combination,
\begin{equation}
\label{eqkn}
K(N)=x_{1}(N)+x_{2}(N),
\end{equation}
as proposed by Kitazawa and Nomura.\cite{kn}

For $j > 0$, the phase boundary approaches that of the isotropic chain as $\Delta$ tends to unity and the logarithmic corrections appear due to the SU(2) symmetry of the problem.  For $j < 0$, this happens only for $\delta > \delta_c$. According to the renormalization group calculation,\cite{kn,gia} these logarithmic corrections are given by,
\begin{equation}
x_{1}(N)=\frac{1}{2}+\frac{3}{4}y_{0}(l)(1+\frac{2}{3}t),
\end{equation}
\begin{equation}
x_{2}(N)=\frac{1}{2}-\frac{1}{4}y_{0}(l)(1+2t),
\end{equation}
\begin{equation}
x_{4}(N)=\frac{1}{2}-\frac{1}{4}y_{0}(l),
\end{equation}
where $x_{4}$ is the scaling dimension of the operator ${\mbox {exp}} (\pm \rm {i}\theta)$, $y_{0}(l)$ is proportional to $1/{\mbox {log}}N$ at the SU(2) symmetric point and the deviation from the SU(2) symmetric point is denoted by $t$.  The scaling dimension $x_{4}$ is calculated by Eq.(\ref{eq3}) for $M^{z}=1$, $P=1$, $k=0$. In this case, we use the combination
\begin{equation}
\label{eqx4}
K(N)+\frac{1}{K(N)}=x_{1}(N)+x_{2}(N)+2x_{4}(N),
\end{equation}
to reduce the influence of the logarithmic corrections for $\Delta \sim 1$.

In both cases, we assume the formula
\begin{equation}
K(N)=K+\frac{C_{1}}{N^{2}}+\frac{C_{2}}{N^{4}},
\end{equation}
to extrapolate $K(N)$ to $N \rightarrow \infty$. The critical exponent $\nu$ of the energy gap is given by
\begin{equation}
\nu=\frac{1}{2-x_1}=\frac{1}{2-\frac{K}{2}},
\end{equation}

For $\Delta \geq 0.9$, we use eq. (\ref{eqx4}) as far as $j > 0$ or $j < 0$ and $\delta > \deltac$. In other cases,  eq. (\ref{eqkn}) is used. The extrapolation procedure of $K$ to $N \rightarrow \infty$ is shown in Fig. \ref{fig3} for $\Delta=0.9, \delta=0.4$ and $j=0.9354$. Figure \ref{fig4} shows the $\Delta$ and $\delta$ dependence of the critical exponent $\nu$. The error bars are estimated from the difference between the values extrapolated  from $12 \leq 4N \leq 20$ and those from $12 \leq 4N \leq 24$. The values of $K$ estimated from other combinations coincide with the above estimation within the error bars. As shown in Fig. \ref{fig4} the critical exponent $\nu$ is almost independent of $\delta$ and decreases with increasing $\Delta$ for $j>0$. At $\Delta=1$ and $0$, the critical exponent $\nu$ reduces to 2/3 and 1, respectively. For $\delta=0$ and $j=1$, the parameter $K$ can be calculated by the Bethe's hypothesis \cite{lp,bik} as
\begin{equation}
K=\frac{\pi}{\pi-\mbox{arccos}\Delta}.
\end{equation}
The numerical results are in agreement with the exact results as shown in Fig. \ref{fig4}.  For $j<0$, the $\Delta$-dependence is rather complicated. For $\mid \delta \mid \ge \deltac$, the phase boundary approaches that of the isotropic model as $\Delta$ tends to unity. At the same time, the critical exponent $\nu$ approaches  2/3 which is the value for the SU(2) symmetric critical point. On the other hand, for $\mid \delta \mid < \deltac$, the phase boundary tends to $j \rightarrow -\infty$ as $\Delta \rightarrow 1$. As discussed in the preceding section, the phase transition in this limit corresponds to the Haldane-large-$D$ transition in the $S=1$ dimerized XXZ chain. Therefore, the critical exponent $\nu$ approaches the nonuniversal value which varies with $\delta$. Although the precise estimation of  $\nu$ for the $S=1$ chain is not available in literature, Glaus and Schneider \cite{gs} estimated as $\nu \simeq 1.5$ at $\delta=0$.  Our results are consistent with their estimation.

Our value of $\nu$ for $\delta = 0$ is also consistent with the critical exponents obtained by Yamanaka, Hatsugai and Kohmoto\cite{yhk} and Yamanaka and Kohmoto\cite{yk}. It should be noted that the sign of exponent $\alpha$ is inverted in Fig. 5 of ref \citen{yhk}\cite{pcy}.

\section{Summary and Discussion}

The ground state phase diagram and the critical exponent $\nu$ of the dimerized and quadrumerized spin-1/2 XXZ chain is calculated by the numerical diagonalization method for $0 < \Delta < 1$. The critical points are determined by the method of twisted boundary condition. For $\delta \simeq 1$ and $j \simeq 0$, the numerical results are compared with those for the effective model obtained perturbationally. The gap exponent $\nu$ is calculated by analyzing the numerical diagonalization data using the conformal field theory.

For $j > 0$, the phase boundaries are insensitive to $\Delta$, while for $j < 0$, they are closed at finite negative values of $j=\jc (\delta)$ and the large-{\it{D}} phase appears in the ground state for $j < \jc (\delta)$.

The $\Delta$-dependence of the exponent $\nu$ is qualitatively different according as $j>0$ or $j<0$. For $j>0$, the critical exponent $\nu$ is insensitive to $\delta$ and always reduces to the SU(2) symmetric value 2/3 as $\Delta \rightarrow 1$. For $j<0$ and $\delta<\deltac$, $\nu$ approaches the nonuniversal value which corresponds to the value for the large-{\it{D}}-Haldane transition in the $S=1$ dimerized XXZ model as $\Delta$ approaches unity. On the other hand, it approaches the SU(2) symmetric value 2/3 at $\Delta=1$ for $j < 0$ and $\delta>\deltac$.

In this paper, we concentrated on the parameter regime $-1 \leq \delta \leq 1$ and $0 \leq \Delta \leq 1$ to clarify the feature of the crossover between the XY case and the isotropic case. It is expected, however, that this model has more variety of phases outside this parameter regime. Especially, from the consideration of the case $\delta=0$\cite{yhk,yk},  we may expect the presence of the XY and ferromagnetic phase for $\Delta < 0$ and  the Ising phase for $\Delta > 1$. It is interesting to find how these phases are modified in the presence of the quadrumerization. The investigation of these problems is left for future studies.

We thank A. Kitazawa, K. Okamoto, S. Yamaguchi and M. Yamanaka for valuable comments and discussion. We are also grateful to H. Nishimori for the program package TITPACK version 2 for the diagonalization of spin-$1/2$ systems. The numerical calculation is performed using the HITAC S820 and SR2201 at the Information Processing Center of Saitama University and the FACOM vpp500 at the Supercomputer Center of Institute for Solid State Physics, University of Tokyo.

\newpage
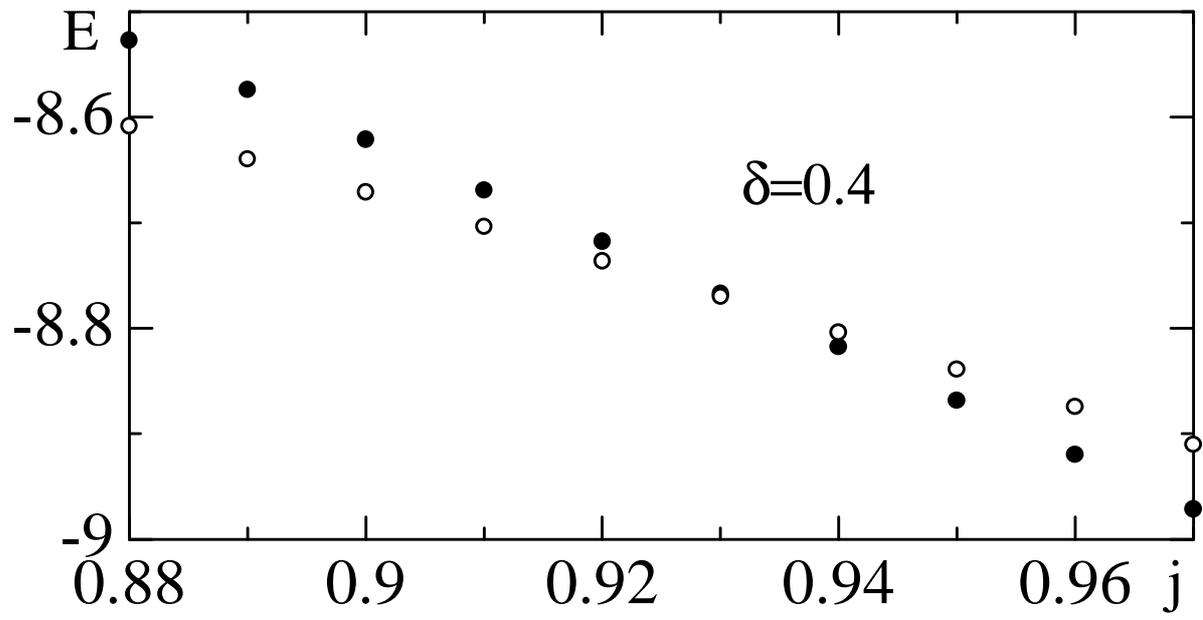
\begin{figure}
\caption{The $j$-dependence of the two lowest energy eigenvalues with twist boundary condition. The energies of the Haldane state and dimer state are represented by $\circ$ and $\bullet$, respectively, for $N=24, \Delta=0.5, \delta=0.4$ and $j>0$. }
\label{fig1}
\end{figure}
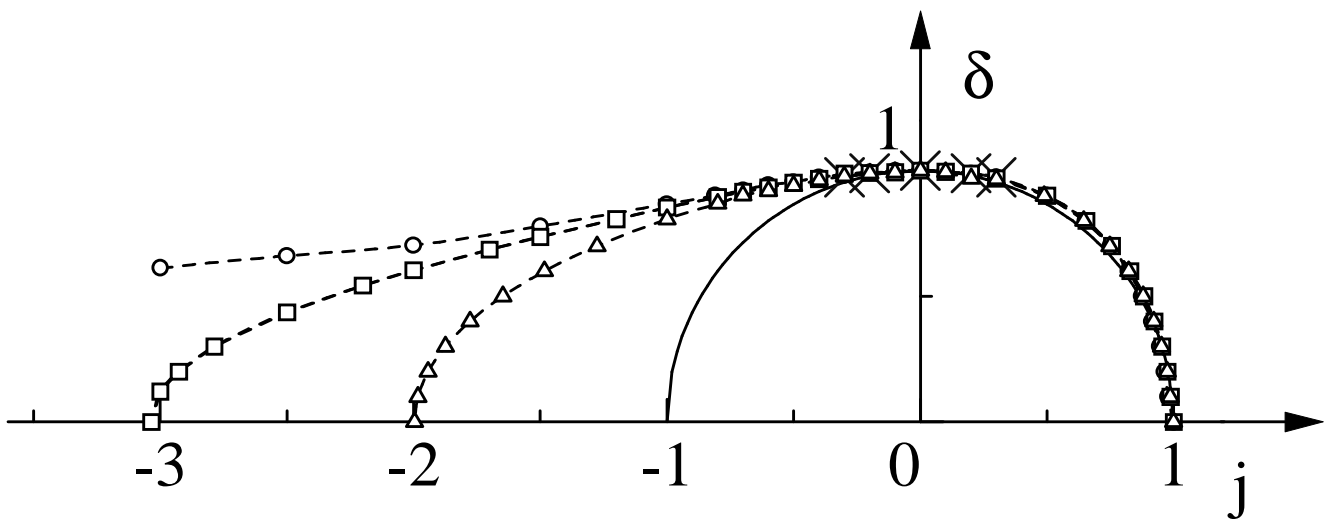
\begin{figure}
\caption{The phase diagram of the isotropic model ($\circ$), $\Delta=0.7$ ($\Box$), $\Delta=0.5$ ($\triangle$) and the XY model (solid line). The result for the effective model (3.3)   is shown by $\times$ for $\Delta=0.5$. The broken lines are the guide for the eye.}
\label{fig2}
\end{figure}
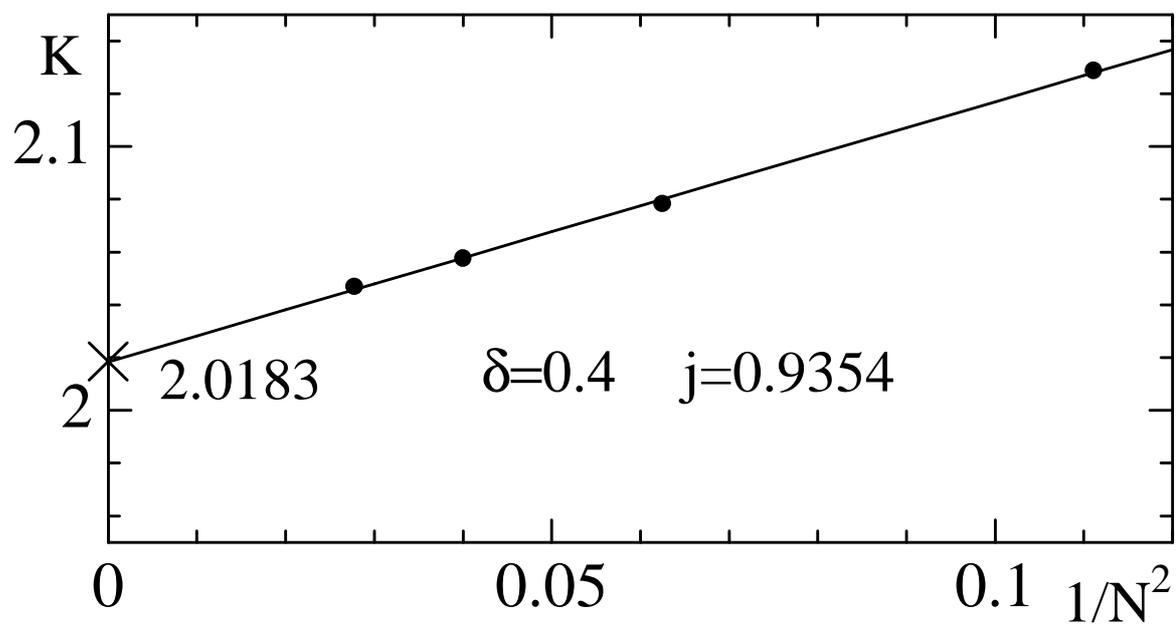
\begin{figure}
\caption{The extrapolation procedure of finite size $K$ for $\Delta=0.9$, $\delta=0.4$ and $j>0$.}
\label{fig3}
\end{figure}
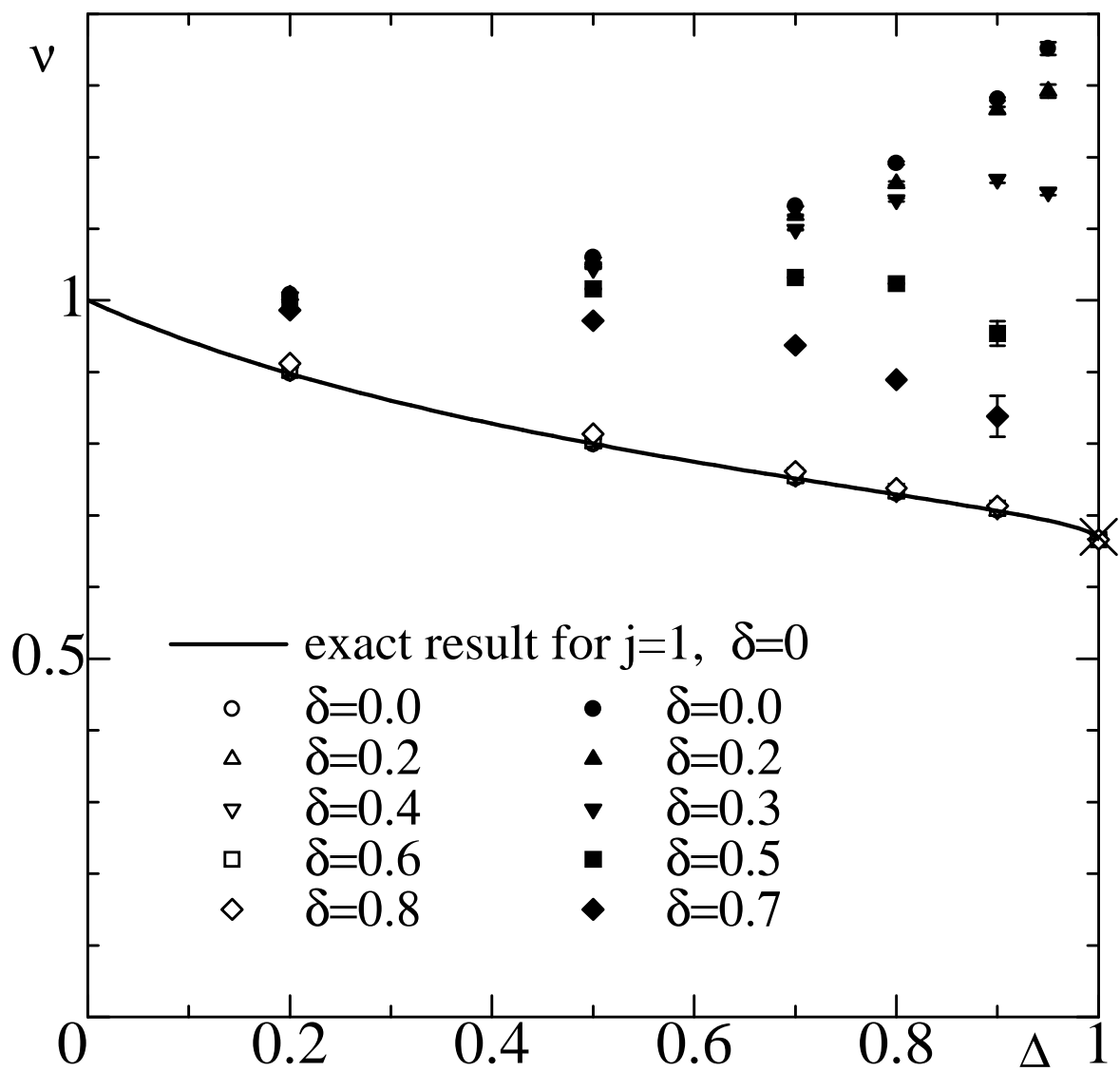
\begin{figure}
\caption{The $\Delta$ dependence of the numerically obtained critical exponent $\nu$. The open and filled symbols represent the cases $j>0$ and $j<0$, respectively. For $j > 0$, the $\delta$-dependence is almost unvisible.}
\label{fig4}
\end{figure}

\begin{thebibliography}{99}
\bibitem{ya}
C. N. Yang and C. P. Yang: Phys. Rev. {\bf 150} (1966) 321.
\bibitem{fd}
F. D. M. Haldane: Phys. Lett. {\bf 93A} (1983) 464; Phys. Rev. Lett. {\bf 50} (1983) 1153.
\bibitem{wk}
W. Chen and K. Hida: J. Phys. Soc. Jpn. {\bf67} (1998) 2910.
\bibitem{ak}
A. Kitazawa and K. Nomura: J. Phys. Soc. Jpn. {\bf 66} (1997) 3379.
\bibitem{kn}
A. Kitazawa and K. Nomura: J. Phys. Soc. Jpn. {\bf 66} (1997) 3944.
\bibitem{tk}
Y.~Kato and A.~Tanaka: J. Phys. Soc. Jpn {\bf 63}
(1994) 1277.
\bibitem{kno}
A. Kitazawa, K. Nomura and K. Okamoto: Phys. Rev. Lett. {\bf 76} (1996) 4038.
\bibitem{tt}
T. Tonegawa, T. Nakao and M. Kaburagi: J. Phys. Soc. Jpn. {\bf 65} (1996) 3317.
\bibitem{ca}
J. L. Cardy: J. Phys. {\bf A17} (1984) L385.
\bibitem{HW}
H. W. Bl\"ote, J. L. Cardy and M. P. Nightingale: Phys. Rev. Lett. {\bf 56} (1986) 742.
\bibitem{IA}
I. Affleck: Phys. Rev. Lett. {\bf 56} (1986) 746; Nucl. Phys. B {\bf 270 [FS16]} (1986) 186.
\bibitem{wkh}
W. Chen, K. Hida and H. Nakano: J. Phys. Soc. Jpn. in press; cond-mat.9808300.
\bibitem{nomura} K. Nomura: J. Phys. A: Math. Gen. {\bf 28} (1995) 5451.
\bibitem{gia}
T. Giamarchi and H. J. Schulz: Phys. Rev. {\bf B39} (1989) 4620.
\bibitem{lp} A. Luther and I. Peschel: Phys. Rev. {\bf B12} (1975) 3908.
\bibitem{bik} N. M. Bogoliubov, A. G. Izergin and V. E. Korepin: Nucl. Phys. {\bf B275} (1986) 687.
\bibitem{gs}
U. Glaus and T. Schneider: Phys. Rev. {\bf B30} (1984) 215.
\bibitem{yhk} M. Yamanaka, Y. Hatsugai and Kohmoto: Phys. Rev. {\bf B50} (1994) 559.
\bibitem{yk} M. Yamanaka and M, Kohmoto: Phys. Rev. {\bf B52} (1995) 1138.
\bibitem{pcy} M. Yamanaka: private communication.


\end{thebibliography}
\end{document}